\documentclass[3p,times]{elsarticle}

\usepackage{ecrc}
\usepackage{amsmath}

\volume{00}

\firstpage{1}

\runauth{}

\jid{procs}

\CopyrightLine{2011}{Published by Elsevier Ltd.}

\usepackage{amssymb}

\usepackage[figuresright]{rotating}

\begin{document}

\begin{frontmatter}



\dochead{}

\title{On the Kullback-Leibler divergence between pairwise isotropic Gaussian-Markov random fields}


\author{Alexandre L. M. Levada}

\address{Computing Department, Federal University of S\~ao Paulo, S\~ao Carlos, SP, Brazil}

\ead{alexandre.levada@ufscar.br}

\begin{abstract}
The Kullback-Leibler divergence or relative entropy is an information-theoretic measure between statistical models that play an important role in measuring a distance between random variables. In the study of complex systems, random fields are mathematical structures that models the interaction between these variables by means of an inverse temperature parameter, responsible for controlling the spatial dependence structure along the field. In this paper, we derive closed-form expressions for the Kullback-Leibler divergence between two pairwise isotropic Gaussian-Markov random fields in both univariate and multivariate cases. The proposed equation allows the development of novel similarity measures in image processing and machine learning applications, such as image denoising and unsupervised metric learning.
\end{abstract}


\end{frontmatter}


\section{Introduction}
\label{sec:intro}
In several research fields, the study of complex systems is critical to understanding the underlying structure of natural phenomena \cite{Holovatch_2017,Rock_2014,CSB,Ross}. Random fields are mathematical structures for modeling non-deterministic behavior in which non-linear interactions between random variables lead to the emergence of long range correlations and phase transitions \cite{Willsky,Phase} and have been employed with success in a wide range of areas, such as statistical mechanics \cite{StatMech}, thermodynamics \cite{Gibbs}, information geometry \cite{Nielsen}, machine learning \cite{GMRF_ML}, image processing \cite{GMRF_Denoising}, among others.  

The objects of interest in this study are pairwise isotropic Gaussian-Markov random fields, that are collections of spatially dependent variables organized in the vertices of a graph, whose set of observable states is continuous \cite{GaussianRandomFields}. The degree of interaction between neighboring variables is quantified by a coupling parameter, also known as the inverse temperature. A natural question that arises in the study of these structures is: how to measure the distance between two Gaussian random fields operating in different regimes? In this paper, we derive closed-form expressions for the Kullback-Leibler divergence between pairwise isotropic Gaussian-Markov random fields, by avoiding the computation of the joint Gibbs distribution by means of the product of the local conditional density functions \cite{MCMA}.

The KL-divergence is a statistical distance, but it is not considered to be a metric in the space of probability distributions. Divergences are usually asymmetric and in certain situations obey a generalized Pythagorean theorem, whereas metrics are symmetric and generalize linear distance, satisfying the triangle inequality \cite{Amari2016,Nielsen}. Relative entropy is directly related to the Fisher information metric. It has been shown that Fisher information is, in fact, the curvature of the relative entropy \cite{Amari2016}. Another important connection between these two information-theoretic measures is that first-order Fisher information is the KL-divergence in the limiting case, that is, when we have two nearby densities $p(x; \vec{\theta})$ and $p(x; \vec{\theta}+\Delta\vec{\theta})$. It has been shown that the KL-divergence between two infinitesimally close densities can be expressed by a quadratic form whose coefficients are given by the elements of the Fisher information matrix \cite{Jeffreys,KL,Kullback_book,Nielsen_Entropy}.  

In order to simplify the mathematical derivation of the KL-divergence between two random field models, we assume some hypothesis: 1) we limit the maximum clique size to two, which is equivalent to having a pairwise interaction random field; 2) the inverse temperature parameter, which controls the spatial dependence structure, is invariant along the random field and isotropic, meaning that the potentials for neighboring interactions are the same in all directions. In practice, the computation of the KL-divergence between two statistical models assumes the assumption that the random variables are independent and identically distributed. However, in random fields, the spatial dependence structure introduces a degree of dependence between the random variables. To the best of our knowledge, closed-form expressions for the KL-divergence between such random field models using local conditional density functions have not been derived yet in the literature.

The remaining of the paper is organized as follows: in Section 2, we derive closed-form expressions for the KL-divergence between univariate pairwise isotropic Gaussian-Markov random fields (GMRF's) and the maximum pseudo-likelihood estimation of the inverse temperature parameter. Section 3 shows the derivation of closed-form expressions for the KL-divergence between multivariate pairwise isotropic Gaussian-Markov random fields and the maximum pseudo-likelihood estimation of the inverse temperature parameter. Finally, Section 4 presents the conclusions and final remarks.

\section{The KL-divergence between univariate pairwise isotropic GMRF's}

\begin{figure}[ht]
\begin{center}
\includegraphics[scale=0.3]{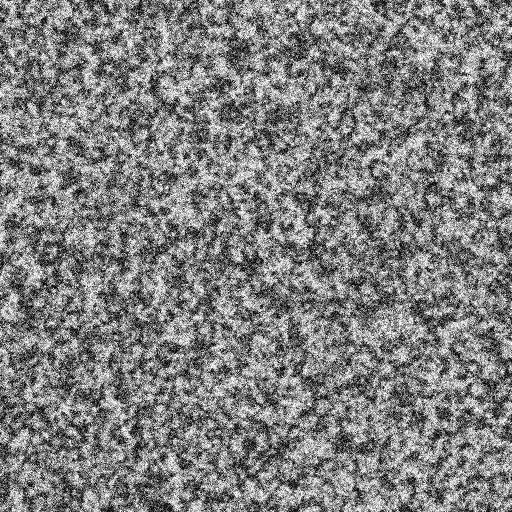}
\includegraphics[scale=0.3]{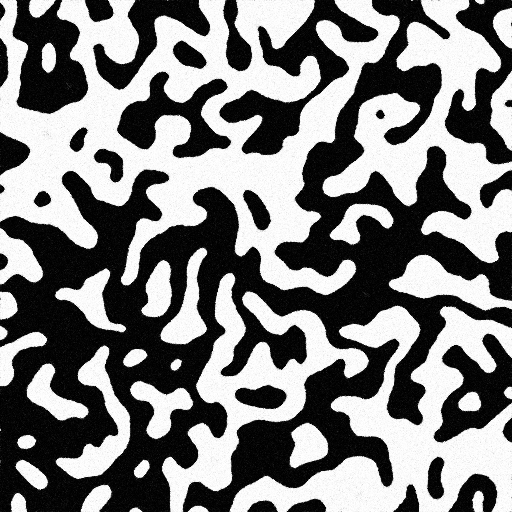}
\end{center}
\caption{Outcomes of a pairwise isotropic Gaussian-Markov random field model.}
\label{fig:GMRF_configs}
\end{figure}

Pairwise isotropy Gaussian-Markov random fields (GMRFs) are mathematical structures that are particularly well suited to the analysis of spatially dependent continuous random variables using non-linear interactions between local regions in a lattice. Compared to other random field models, the major advantage of GMRF's is the mathematical tractability. We are frequently unable to obtain closed-form expressions for certain information-theoretic metrics because they need the computation of exact expectations. It is possible to compute these quantities in pairwise isotropic GMRFs without using numerical techniques, which greatly decreases the computational cost. The exact closed-form equations for the components of the Fisher information matrix are a good example \cite{MCMA}. Furthermore, we characterize a pairwise isotropic Gaussian-Markov random field by a collection of local conditional density functions by invoking the Hammersley-Clifford theorem \cite{Hammersley}, which establishes the equivalence between Gibbs random fields (global models) and Markov random fields (local models). The local conditional density function of a pairwise isotropic GMRF model is given by:

\begin{equation}
	p\left( x_{i} | \eta_{i}, \vec{\theta} \right) = \frac{1}{\sqrt{2\pi\sigma^2}}exp\left\{-\frac{1}{2\sigma^{2}} \left[ x_{i} - \mu - \beta \sum_{j \in \eta_{i}} \left( x_{j} - \mu \right) \right]^{2} \right\} \qquad i=1,...,n
	\label{eq:GMRF}
\end{equation} where $\eta_i$ represents the neighborhood around the i-ht variable (usually it is a second-order neighborhood system comprised by the 8 nearest neighbors of $x_i$), $\vec{\theta} = (\mu, \sigma^{2}, \beta)$ is the vector of model parameters, with $\mu$ and $\sigma^{2}$ being, the expected value (mean) and the variance of the random variables in the lattice, and $\beta$ being the inverse temperature, which encodes the spatial dependence between the variables in the field. Note that when $\beta = 0$, the random field model degenerates to a regular Gaussian distribution (the variables become statistically independent). The main advantage of using the local model is that we avoid the joint Gibbs distribution, which is computationally intractable. Figure \ref{fig:GMRF_configs} shows two outcomes of a pairwise isotropic Gaussian-Markov random field along a Markov Chain Monte Carlo simulation.

\subsection{The Kullback-Leibler divergence}

The Kullback-Leibler divergence is the relative entropy, that is, the difference between the cross-entropy of $p(x)$ and $q(x)$ and the entropy of $p(x)$:

\begin{align}
	D_{KL}(p, q) = H(p, q) - H(p) = \int p(x) log \left(\frac{p(x)}{q(x)}\right) dx = E_{p} \left[ log \left( \frac{p(x)}{q(x)} \right) \right]
\end{align}

It should be mentioned that the relative entropy is always non-negative, that is, $D_{KL}(p, q) \geq 0$, being equal to zero if, and only if, $p(x) = q(x)$. The KL-divergence between two pairwise isotropic GMRF's with local conditional density functions $p(x_i | \eta_i, \vec{\theta}_1)$ and $q(x_i | \eta_i, \vec{\theta}_2)$, for $i=1,2,...,n$, with parameter vectors $\vec{\theta}_1 = (\mu_1, \sigma_1^2, \beta_1)$ and $\vec{\theta}_2 = (\mu_2, \sigma_2^2, \beta_2)$, is given by:

\begin{align}
	\label{eq:DKLpq}
	D_{KL}(p, q) = log \left( \frac{\sigma_2}{\sigma_1} \right) & + \frac{1}{2\sigma_2^2} E_p\left\{ \left[ x - \mu_2 - \beta_2 \sum_{j \in \eta_i} (x_j - \mu_2) \right]^2 \right\} - \frac{1}{2\sigma_1^2} E_p\left\{ \left[ x - \mu_1 - \beta_1 \sum_{j \in \eta_i} (x_j - \mu_1) \right]^2 \right\}
\end{align}

It is possible to expand the second expected value in equation \eqref{eq:DKLpq} as:

\begin{align}
	& E_p\left\{ \left[ x - \mu_1 - \beta_1 \sum_{j \in \eta_i} (x_j - \mu_1) \right]^2 \right\} = E_p\left\{ \left( x - \mu_1 \right)^2 - 2\beta_1 \sum_{j \in \eta_i}(x_i - \mu_1) (x_j - \mu_1) + \beta_1^2 \sum_{j \in \eta_i}\sum_{k \in \eta_i}(x_j - \mu_1) (x_k - \mu_1) \right\} \\ \nonumber & = \sigma_1^2 - 2 \beta_1 \sum_{j \in \eta_i} \sigma_{ij} + \beta_1^2 \sum_{j \in \eta_i}\sum_{k \in \eta_i} \sigma_{jk}
\end{align} where $\sigma_{ij}$ denotes the covariance between the central variable $x_i$ and a neighboring variable $x_j$ and $\sigma_{jk}$ denotes the covariance between two different variables belonging to the same neighborhood system $\eta_i$. Now, we expand the first expected value as:

\begin{align}
	\label{eq:Ep2}
	& E_p\left\{ \left[ x - \mu_2 - \beta_2 \sum_{j \in \eta_i} (x_j - \mu_2) \right]^2 \right\} = E_p\left\{ \left( x - \mu_2 \right)^2 - 2\beta_2 \sum_{j \in \eta_i}(x_i - \mu_2) (x_j - \mu_2) + \beta_2^2 \sum_{j \in \eta_i}\sum_{k \in \eta_i}(x_j - \mu_2) (x_k - \mu_2) \right\}
\end{align}

We need to break equation \eqref{eq:Ep2} in three distinct expected values in order to simplify it. The first one remains identical to the corresponding term in the univariate Gaussian case:

\begin{align}
	E_p\left\{ \left( x - \mu_2 \right)^2 \right\} & = E_p[x^2] - 2E_p[x]\mu_2 + \mu_2^2 = \sigma_1^2 + \mu_1^2 - 2E_p[x]\mu_2 + \mu_2^2 = \sigma_1^2 + (\mu_1 - \mu_2)^2
\end{align}

The second expected value is given by:

\begin{align}
- 2\beta_2 \sum_{j \in \eta_i} E_p\left\{ (x_i - \mu_2) (x_j - \mu_2) \right\} = - 2\beta_2 \sum_{j \in \eta_i} E_p\left\{ x_i x_j - \mu_2 x_i - \mu_2 x_j + \mu_2^2 \right\} = - 2\beta_2 \sum_{j \in \eta_i} \left[ \sigma_{ij}^{(p)} + (\mu_1 - \mu_2)^2 \right]
\end{align} where $\sigma_{ij}^{(p)}$ denotes the covariance between $x_i$ and $x_j$. In a similar way, the third expected value leads to:

\begin{align}
	\beta_2^2 \sum_{j \in \eta_i}\sum_{k \in \eta_i} E_p\left\{ (x_j - \mu_2) (x_k - \mu_2) \right\} = \beta_2^2 \sum_{j \in \eta_i}\sum_{k \in \eta_i} \left[ \sigma_{jk}^{(p)} + (\mu_1 - \mu_2)^2 \right]
\end{align}

Thus, the KL-divergence can be expressed as:

\begin{align}
	D_{KL}(p, q) = log \left( \frac{\sigma_2}{\sigma_1} \right) - \frac{1}{2\sigma_1^2} \left[ \sigma_1^2 - 2 \beta_1 \sum_{j \in \eta_i} \sigma_{ij} + \beta_1^2 \sum_{j \in \eta_i}\sum_{k \in \eta_i} \sigma_{jk} \right] & + \frac{1}{2\sigma_2^2} \left\{ \sigma_1^2 + (\mu_1 - \mu_2)^2 - 2\beta_2 \sum_{j \in \eta_i} \left[ \sigma_{ij}^{(p)} + (\mu_1 - \mu_2)^2 \right] \right. \nonumber \\  & \left. \qquad\qquad + \beta_2^2 \sum_{j \in \eta_i}\sum_{k \in \eta_i} \left[ \sigma_{jk}^{(p)} + (\mu_1 - \mu_2)^2 \right] \right\}
\end{align}

Note that the previous equation is equivalent to:

\begin{align}
	& D_{KL}(p, q) = log \left( \frac{\sigma_2}{\sigma_1} \right) - \frac{1}{2\sigma_1^2} \left[ \sigma_1^2 - 2 \beta_1 \sum_{j \in \eta_i} \sigma_{ij}^{(p)} + \beta_1^2 \sum_{j \in \eta_i}\sum_{k \in \eta_i} \sigma_{jk}^{(p)} \right] \\ \nonumber & + \frac{1}{2\sigma_2^2} \left\{ \left[ \sigma_1^2 - 2 \beta_2 \sum_{j \in \eta_i} \sigma_{ij}^{(p)} + \beta_2^2 \sum_{j \in \eta_i}\sum_{k \in \eta_i} \sigma_{jk}^{(p)} \right]   + \left[ (\mu_1 - \mu_2)^2 - 2 \Delta \beta_2 (\mu_1 - \mu_2)^2 + \Delta^2 \beta^2 (\mu_1 - \mu_2)^2 \right] \right\}
\end{align} where $\Delta$ denotes the number of elements in the neighborhood system $\eta_i$. By simplifying the last term of the previous equation, we have the final expression for the KL-divergence:

\begin{align}
	& D_{KL}(p, q) = log \left( \frac{\sigma_2}{\sigma_1} \right) - \frac{1}{2\sigma_1^2} \left[ \sigma_1^2 - 2 \beta_1 \sum_{j \in \eta_i} \sigma_{ij}^{(p)} + \beta_1^2 \sum_{j \in \eta_i}\sum_{k \in \eta_i} \sigma_{jk}^{(p)} \right] \\ \nonumber & + \frac{1}{2\sigma_2^2} \left\{ \left[ \sigma_1^2 - 2 \beta_2 \sum_{j \in \eta_i} \sigma_{ij}^{(p)} + \beta_2^2 \sum_{j \in \eta_i}\sum_{k \in \eta_i} \sigma_{jk}^{(p)} \right] + (\mu_1 - \mu_2)^2 (1 - \Delta \beta_2)^2 \right\}
\end{align}  

Note that, when the inverse temperatures are zero, that is, $\beta_1 = \beta_2 = 0$, $D_{KL}(p, q)$ becomes the KL-divergence between two Gaussian pdf's. By analogy, the expression for $D_{KL}(q, p)$ can be written as:

\begin{align}
	& D_{KL}(q, p) = log \left( \frac{\sigma_1}{\sigma_2} \right) - \frac{1}{2\sigma_2^2} \left[ \sigma_2^2 - 2 \beta_2 \sum_{j \in \eta_i} \sigma_{ij}^{(q)} + \beta_2^2 \sum_{j \in \eta_i}\sum_{k \in \eta_i} \sigma_{jk}^{(q)} \right] \\ \nonumber & + \frac{1}{2\sigma_1^2} \left\{ \left[ \sigma_2^2 - 2 \beta_1 \sum_{j \in \eta_i} \sigma_{ij}^{(q)} + \beta_1^2 \sum_{j \in \eta_i}\sum_{k \in \eta_i} \sigma_{jk}^{(q)} \right] + (\mu_1 - \mu_2)^2 (1 - \Delta \beta_1)^2 \right\}
\end{align}

The symmetrized KL-divergence is obtained by averaging $D_{KL}(p, q)$ and $D_{KL}(q, p)$. For pairwise isotropic GMRF's models, it is given by:

\begin{align}
	D_{KL}^{sym}(p, q) & = \frac{1}{4\sigma_1^2\sigma_2^2} \Bigg\{ \left( \sigma_1^2 - \sigma_2^2 \right)^2 - 2(\beta_2\sigma_1^2 - \beta_1\sigma_2^2)\sum_{j \in \eta_i}\left(\sigma_{ij}^{(p)} - \sigma_{ij}^{(q)}\right) + (\beta_2^2\sigma_1^2 - \beta_1^2\sigma_2^2)\sum_{j \in \eta_i}\sum_{k \in \eta_i}\left(\sigma_{jk}^{(p)} - \sigma_{jk}^{(q)}\right) \\ \nonumber & \qquad\qquad\qquad\qquad\qquad\qquad + \left( \mu_1 - \mu_2 \right)^2 \left[ \sigma_1^2 (1 - \Delta\beta_2)^2 + \sigma_2^2 (1 - \Delta\beta_1)^2 \right] \Bigg\}
\end{align}

\begin{figure}[ht]
\begin{center}
\includegraphics[scale=0.4]{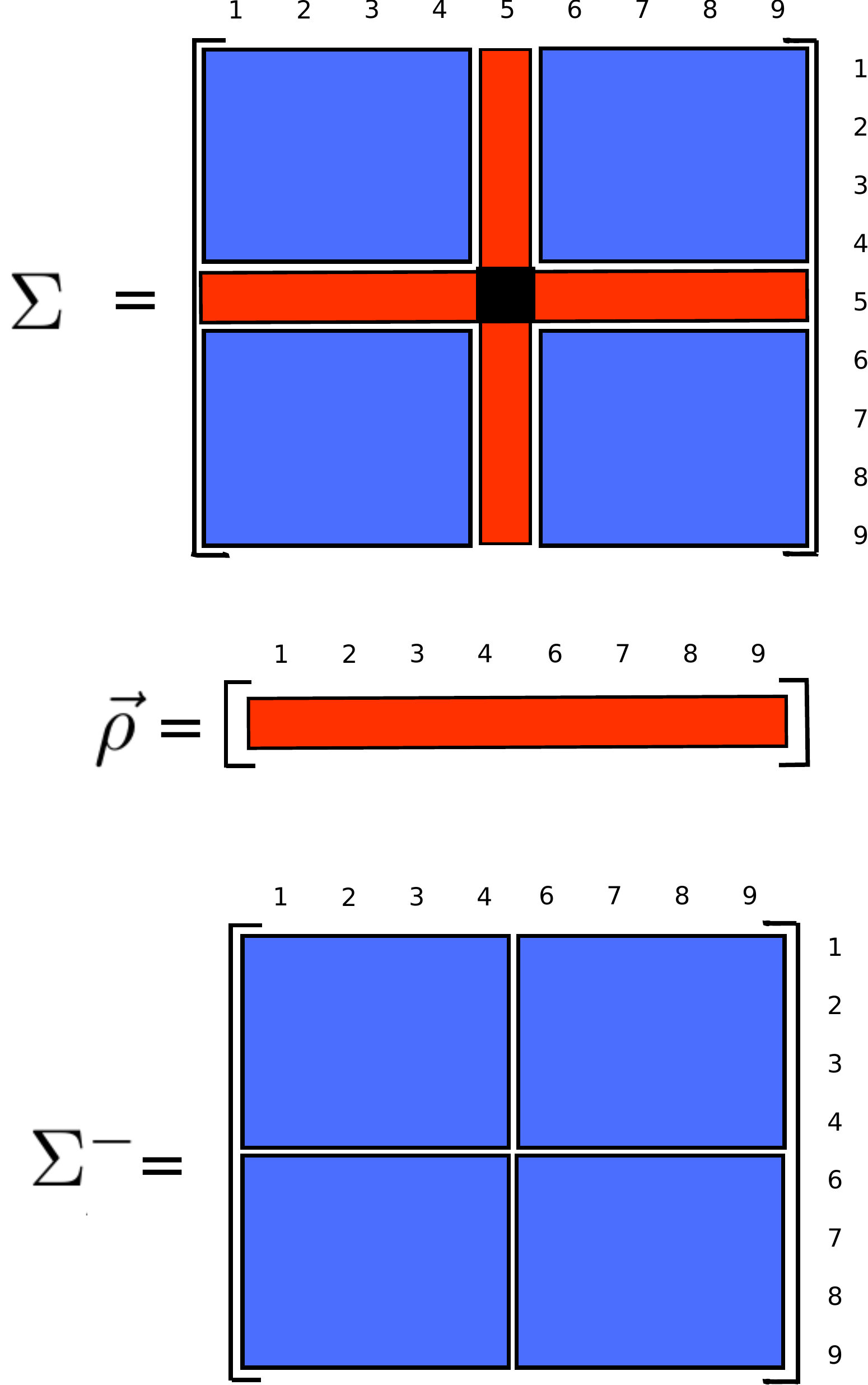}
\end{center}
\caption{Decomposition of $\Sigma_{p}$ into $\Sigma_{p}^{-}$ and $\vec{\rho}$ on a second-order neighborhood system ($\Delta=8$). By rewriting the components of the Fisher information matrices in terms of Kronocker products, we speed up the computational simulations.
}
\label{fig:cov_matrix}
\end{figure}

It is possible to speed up the computation of the KL-divergence by using an alternative representation. First, note that we can convert each $3 \times 3$ neighborhood patch formed by $x_{i} \cup \eta_{i}$ into a vector $p_i$ of 9 elements by piling its rows. Then, we compute the covariance matrix of these vectors, for $i = 1, 2,..., n$ denoted by $\Sigma_{p}$. From this covariance matrix, we extract two main components: 1) a vector of size 8, $\vec{\rho}$, composed by the the elements of the central row of $\Sigma_{p}$, excluding the middle one, which denotes the variance of $x_i$ (we want only the covariances between $x_i$ and $x_j$, for $j \neq i$; and 2) a sub-matrix of dimensions $8 \times 8$, $\Sigma_{p}^{-}$, obtained by removing the central row and central column from $\Sigma_{p}$ (we want only the covariances between $x_j \in \eta_i$ and $x_k \in \eta_i$). Figure \ref{fig:cov_matrix} shows the decomposition of the covariance matrix $\Sigma_{p}$ into the sub-matrix $\Sigma_{p}^{-}$ and the vector $\vec{\rho}$.

We can now express the KL-divergence as:

\begin{align}
	D_{KL}(p, q) & = log \left( \frac{\sigma_2}{\sigma_1} \right) - \frac{1}{2\sigma_1^2} \left[ \sigma_1^2 - 2\beta_1 \left\| \vec{\rho}^{(p)} \right\|_{+} + \beta_1^2 \left\| \Sigma_{(p)}^{-} \right\|_{+} \right] \\ \nonumber & + \frac{1}{2\sigma_2^2} \left\{ \left[ \sigma_1^2 - 2\beta_2 \left\| \vec{\rho}^{(p)} \right\|_{+} + \beta_2^2 \left\| \Sigma_{(p)}^{-} \right\|_{+}  \right] + (\mu_1 - \mu_2)^2 (1 - \Delta\beta_2)^2 \right\}   
	\label{eq:KL_GMRFA} 
\end{align} where $\left\| A \right\|_{+}$ denotes the summation of the elements of the matrix/vector $A$. Finally, the symmetrized KL-divergence can be expressed by:

\begin{align}
	D_{KL}^{sym}(p, q) & = \frac{1}{4\sigma_1^2\sigma_2^2} \Bigg\{ \left( \sigma_1^2 - \sigma_2^2 \right)^2 - 2(\beta_2\sigma_1^2 - \beta_1\sigma_2^2)\left( \left\| \vec{\rho}^{(p)} \right\|_{+} - \left\| \vec{\rho}^{(q)} \right\|_{+} \right) \\ \nonumber & + (\beta_2^2\sigma_1^2 - \beta_1^2\sigma_2^2)\left( \left\| \Sigma_{(p)}^{-} \right\|_{+} - \left\| \Sigma_{(q)}^{-} \right\|_{+}  \right) + \left( \mu_1 - \mu_2 \right)^2 \left[ \sigma_1^2 (1 - \Delta\beta_2)^2 + \sigma_2^2 (1 - \Delta\beta_1)^2 \right] \Bigg\}
	\label{eq:KL_GMRF_sym} 
\end{align} where $\vec{\rho}^{(p)}$ and $\Sigma_{(p)}^{-}$ are extracted from the covariance matrix of the first GMRF and $\vec{\rho}^{(q)}$ and $\Sigma_{(q)}^{-}$ are extracted from the covariance matrix of the second GMRF.

\subsection{Maximum pseudo-likelihood estimation of the inverse temperature}

The KL-divergence between pairwise isotropic GMRF's is function of the model parameters, more precisely, of the variance, covariances and the inverse temperature. It has been shown that maximum pseudo-likelihood estimation of the inverse temperature in random fields is a viable option because it avoids the joint Gibbs distribution \cite{Besag}. Both maximum likelihood (ML) and maximum pseudo-likelihood (MPL) estimators are asymptotically normal \cite{MPL2,MPL}, however, unbiasedness is not granted by either ML or MPL estimation. Actually, there is no method that guarantees the existence of unbiased estimators for a fixed size sample \cite{Bickel,Lehman,Casella}. However, while ML estimators are asymptotic unbiased and efficient, that is, their asymptotic variance reaches the Cramér-Rao lower bound \cite{Cramer,Rao}, there is no result  proving that the same is valid for MPL estimators. In the isotropic pairwise Gaussian-Markov random field model, we have to maximize the following objective function (pseudo-likelihood function):

\begin{equation}
	log~L\left(\vec{\theta}; \mathbf{X} \right) = -\frac{n}{2}log\left( 2\pi\sigma^{2} \right) -\frac{1}{2\sigma^{2}}\sum_{i=1}^{n}\left[ x_{i} - \mu - \beta\sum_{j \in \eta_i}\left( x_{j} - \mu \right) \right]^{2}
	\label{eq:GMRF_PL} 
\end{equation}

If differentiate equation \eqref{eq:GMRF_PL} with respect to $\beta$ and properly solve the pseudo-likelihood equation, we have:

\begin{equation}
	\hat{\beta}_{MPL} = \frac{\displaystyle\sum_{i=1}^{n}\left[\left( x_{i} - \mu \right)\displaystyle\sum_{j \in \eta_i}\left(x_{j} - \mu \right)\right]}{\displaystyle\sum_{i=1}^{n}\left[ \displaystyle\sum_{j \in \eta_i}\left( x_{j} - \mu  \right) \right]^{2}}
	\label{eq:BetaMPL}
\end{equation}

Another simplifying assumption is that in our case, the random field is defined on a 2D lattice, where the size of the neighborhood system is fixed ($\Delta = 8$). Expanding the square in the denominator, dividing both the numerator and the denominator by $n$ (sample size) and exchanging the order of the summations, we have:

\begin{equation}
	\hat{\beta}_{MPL} = \frac{\displaystyle \sum_{j \in \eta_i} \left[\displaystyle \frac{1}{n}\sum_{i=1}^{n} \left( x_{i} - \mu \right) \left(x_{j} - \mu \right)\right]}{\displaystyle \sum_{j \in \eta_i}\sum_{k \in \eta_i} \left[\displaystyle \frac{1}{n}\sum_{i=1}^{n} \left( x_{j} - \mu \right) \left(x_{k} - \mu \right)\right]} = \frac{\displaystyle\sum_{j \in \eta_i}{\sigma}_{ij}}{\displaystyle\sum_{j \in \eta_i}\displaystyle\sum_{k \in \eta_i}{\sigma}_{jk}}  = \frac{\left\| \vec{\rho} \right\|_{+}}{\left\| \Sigma_{p}^{-}\right\|_{+}}
\end{equation} which means that we can use the sample covariance matrix of the vectors extracted from the configuration patterns. As a consequence, a sequence of Gaussian random field outcomes can be summarized into a sequence of covariance matrices. In computational terms, it means a huge reduction in the volume of data. 



\section{The KL-divergence between multivariate pairwise isotropic GMRF's}

The local conditional density functions of the univariate model can be easily generalized for the multivariate case:

\begin{align}
	p \left( \vec{x}_i | \eta_{i}, \vec{\theta}  \right) = \frac{1}{(2\pi)^{\frac{d}{2}} \lvert \Sigma \rvert^{\frac{1}{2}}} exp \left\{ -\frac{1}{2} \left[ \vec{x}_i - \vec{\mu} - \beta \sum_{j \in \eta} \left( \vec{x}_j - \vec{\mu} \right)\right]^T \Sigma^{-1} \left[ \vec{x}_i - \vec{\mu} - \beta \sum_{j \in \eta} \left( \vec{x}_j - \vec{\mu} \right)\right] \right\}
\end{align} where $\vec{\theta} = \left( \vec{\mu}, \Sigma, \beta \right)$ is the vector of parameters, with $\vec{\mu}$ denoting the mean vector, $\Sigma$ denoting the covariance matrix and $\beta$ the inverse temperature. Then, the KL-divergence between two multivariate pairwise isotropic GMRF's with local conditional density functions $p(\vec{x}_i | \eta_i, \vec{\theta}_1)$ and $q(\vec{x}_i | \eta_i, \vec{\theta}_2)$, for $i=1,2,...,n$, with parameter vectors $\vec{\theta}_1 = (\vec{\mu}_1, \Sigma_1, \beta_1)$ and $\vec{\theta}_2 = (\vec{\mu}_2, \Sigma_2, \beta_2)$, is given by:

\begin{align}
	D_{KL}(p, q) = E_p \left[ log \frac{p(\vec{x})}{q(\vec{x})} \right] = \frac{1}{2} log \frac{\lvert \Sigma_2 \rvert}{\lvert \Sigma_1 \rvert} & - \frac{1}{2} E_p \left\{ \left[ \vec{x}_i - \vec{\mu}_1 - \beta_1 \sum_{j \in \eta} \left( \vec{x}_j - \vec{\mu}_1 \right)\right]^T \Sigma_1^{-1} \left[ \vec{x}_i - \vec{\mu}_1 - \beta_1 \sum_{j \in \eta} \left( \vec{x}_j - \vec{\mu}_1 \right)\right] \right\} \\ \nonumber & + \frac{1}{2} E_p \left\{ \left[ \vec{x}_i - \vec{\mu}_2 - \beta_2 \sum_{j \in \eta} \left( \vec{x}_j - \vec{\mu}_2 \right)\right]^T \Sigma_2^{-1} \left[ \vec{x}_i - \vec{\mu}_2 - \beta_2 \sum_{j \in \eta} \left( \vec{x}_j - \vec{\mu}_2 \right)\right] \right\}
	\label{eq:DKL}
\end{align}

We will call the first expected value A, and the second one, B. Then, by direct application of the distributive law, we have:

\begin{align}
	A = E_p\left[ \left( \vec{x}_i - \vec{\mu}_1 \right)^T \Sigma_1^{-1} \left( \vec{x}_i - \vec{\mu}_1 \right) \right] & - 2\beta_1 \sum_{j\in\eta_i} E_p \left[ \left( \vec{x}_i - \vec{\mu}_1 \right)^T \Sigma_1^{-1} \left( \vec{x}_j - \vec{\mu}_1 \right) \right] + \beta_1^2 \sum_{j\in\eta_i}\sum_{k\in\eta_i} E_p \left[ \left( \vec{x}_j - \vec{\mu}_1 \right)^T \Sigma_1^{-1} \left( \vec{x}_k - \vec{\mu}_1 \right) \right]
\end{align} where we will call the first expected value $A_1$, the second one $A_2$ and the third one $A_3$. Hence, knowing that the argument of the expectation in $A_1$ is a scalar, we have:

\begin{align}
	A_1 = E_p\left[ \left( \vec{x}_i - \vec{\mu}_1 \right)^T \Sigma_1^{-1} \left( \vec{x}_i - \vec{\mu}_1 \right) \right] = E_p\left\{ Tr \left[ \left( \vec{x}_i - \vec{\mu}_1 \right)^T \Sigma_1^{-1} \left( \vec{x}_i - \vec{\mu}_1 \right) \right] \right\}
\end{align} where $Tr(.)$ denotes the trace operator. As the trace operator is invariant under cyclic permutations, we can write:

\begin{align}
	A_1 = E_p\left\{ Tr \left[ \Sigma_1^{-1} \left( \vec{x}_i - \vec{\mu}_1 \right)  \left( \vec{x}_i - \vec{\mu}_1 \right)^T \right] \right\} 
\end{align}

By the linearity of the expected value, we have:

\begin{align}
	A_1 = Tr\left\{ \Sigma_1^{-1} E_p \left[ \left( \vec{x}_i - \vec{\mu}_1 \right)  \left( \vec{x}_i - \vec{\mu}_1 \right)^T \right] \right\} = Tr\left[ \Sigma_1^{-1} \Sigma_1 \right] = Tr \left[ I_d \right] = d
\end{align}

Moving forward to the second term ($A_2$), by using a similar deduction, we can write:

\begin{align}
	A_2 & = E_p\left[ \left( \vec{x}_i - \vec{\mu}_1 \right)^T \Sigma_1^{-1} \left( \vec{x}_j - \vec{\mu}_1 \right) \right] = Tr\left\{ \Sigma_1^{-1} E_p \left[ \left( \vec{x}_i - \vec{\mu}_1 \right)  \left( \vec{x}_j - \vec{\mu}_1 \right)^T \right] \right\} = Tr \left[ \Sigma_1^{-1} \Sigma_1^{(ij)} \right]
\end{align} where $\Sigma_1^{(ij)}$ is the cross-covariance matrix between the random vectors $\vec{x}_i$ and $\vec{x}_j$ computed in the random field with vector of parameters $\vec{\theta}_1 = (\vec{\mu}_1, \Sigma_1, \beta_1)$. Similarly, the third term ($A_3$) is given by:

\begin{align}
	A_3 & = E_p\left[ \left( \vec{x}_j - \vec{\mu}_1 \right)^T \Sigma_1^{-1} \left( \vec{x}_k - \vec{\mu}_1 \right) \right] = Tr\left\{ \Sigma_1^{-1} E_p \left[ \left( \vec{x}_j - \vec{\mu}_1 \right)  \left( \vec{x}_k - \vec{\mu}_1 \right)^T \right] \right\} = Tr \left[ \Sigma_1^{-1} \Sigma_1^{(jk)} \right]
\end{align} where $\Sigma_1^{(jk)}$ is the cross-covariance matrix between the random vectors $\vec{x}_j$ and $\vec{x}_k$ belonging to the same neighborhood system and computed in the random field with vector of parameters $\vec{\theta}_1 = (\vec{\mu}_1, \Sigma_1, \beta_1)$. Hence, the final expression for $A$ becomes:

\begin{align}
	A = -\frac{1}{2} \left\{ d - 2\beta_1 \sum_{j\in\eta_i} Tr \left[ \Sigma_1^{-1} \Sigma_1^{(ij)} \right] + \beta_1^2 \sum_{j\in\eta_i} \sum_{k\in\eta_i} Tr \left[ \Sigma_1^{-1} \Sigma_1^{(jk)} \right] \right\}
\end{align}

We now proceed with the expansion of the second large expectation in equation \eqref{eq:DKL}, which we called $B$. Note that, by direct application of the distributive law, $B$ can also be split into three distinct terms as:

\begin{align}
	B = E_p\left[ \left( \vec{x}_i - \vec{\mu}_2 \right)^T \Sigma_2^{-1} \left( \vec{x}_i - \vec{\mu}_2 \right) \right] & - 2\beta_2 \sum_{j\in\eta_i} E_p \left[ \left( \vec{x}_i - \vec{\mu}_2 \right)^T \Sigma_2^{-1} \left( \vec{x}_j - \vec{\mu}_2 \right) \right] + \beta_2^2 \sum_{j\in\eta_i}\sum_{k\in\eta_i} E_p \left[ \left( \vec{x}_j - \vec{\mu}_2 \right)^T \Sigma_2^{-1} \left( \vec{x}_k - \vec{\mu}_2 \right) \right]
\end{align} where the first term is called $B_1$, the second one $B_2$ and the third one $B_3$. By expanding the first term ($B_1$) using the same steps applied in the previous equations, we have:

\begin{align}
	B_1 = Tr\left\{ \Sigma_2^{-1} E_p \left[ \left( \vec{x}_i - \vec{\mu}_2 \right)  \left( \vec{x}_i - \vec{\mu}_2 \right)^T \right] \right\}
\end{align}

Since the expectation involves the local conditional density function $p(\vec{x})$, we have to expand the quadratic form, leading to:

\begin{align}
	B_1 & = Tr\left\{ \Sigma_2^{-1} E_p \left[ \vec{x}_i \vec{x}_i^T - 2 \vec{x}_i \vec{\mu}_2^T  + \vec{\mu}_2 \vec{\mu}_2^T \right] \right\} = Tr\left\{ \Sigma_2^{-1} \left[ \Sigma_1 + \vec{\mu}_1\vec{\mu}_1^T - 2\vec{\mu}_1\vec{\mu}_2^T + \vec{\mu}_2\vec{\mu}_2^T \right] \right\} \\ \nonumber & = Tr \left[ \Sigma_2^{-1} \Sigma_1 \right] + Tr \left[ \Sigma_2^{-1} \left( \vec{\mu}_1 - \vec{\mu}_2 \right) \left( \vec{\mu}_1 - \vec{\mu}_2 \right)^T \right] = Tr \left[ \Sigma_2^{-1} \Sigma_1 \right] + \left( \vec{\mu}_1 - \vec{\mu}_2 \right)^T \Sigma_2^{-1} \left( \vec{\mu}_1 - \vec{\mu}_2 \right)
\end{align} where the second term is the Mahalanobis distance between the means. Using a similar deduction, the second term of $B$ can be expressed as:

\begin{align}
	B_2 & = Tr\left\{ \Sigma_2^{-1} E_p \left[ \left( \vec{x}_i - \vec{\mu}_2 \right)  \left( \vec{x}_j - \vec{\mu}_2 \right)^T \right] \right\} = Tr\left\{ \Sigma_2^{-1} E_p \left[ \vec{x}_i \vec{x}_j^T - \vec{x}_i \vec{\mu}_2^T - \vec{x}_j \vec{\mu}_2^T  + \vec{\mu}_2 \vec{\mu}_2^T \right] \right\} \\ \nonumber & = Tr\left\{ \Sigma_2^{-1} \left[ \Sigma_1^{(ij)} + \vec{\mu}_1\vec{\mu}_1^T - 2\vec{\mu}_1\vec{\mu}_2^T + \vec{\mu}_2\vec{\mu}_2^T \right] \right\} = Tr \left[ \Sigma_2^{-1} \Sigma_1^{(ij)} \right] + Tr \left[ \Sigma_2^{-1} \left( \vec{\mu}_1 - \vec{\mu}_2 \right) \left( \vec{\mu}_1 - \vec{\mu}_2 \right)^T \right] \\ \nonumber & = Tr \left[ \Sigma_2^{-1} \Sigma_1^{(ij)} \right] + \left( \vec{\mu}_1 - \vec{\mu}_2 \right)^T \Sigma_2^{-1} \left( \vec{\mu}_1 - \vec{\mu}_2 \right)
\end{align} where $\Sigma_1^{(ij)}$ is the cross-covariance matrix between the random vectors $\vec{x}_i$ and $\vec{x}_j$ computed in the random field with vector of parameters $\vec{\theta}_1 = (\vec{\mu}_1, \Sigma_1, \beta_1)$. Finally, the third term of $B$ is given by:

\begin{align}
	B_3 & = Tr\left\{ \Sigma_2^{-1} E_p \left[ \left( \vec{x}_j - \vec{\mu}_2 \right)  \left( \vec{x}_k - \vec{\mu}_2 \right)^T \right] \right\} = Tr\left\{ \Sigma_2^{-1} E_p \left[ \vec{x}_j \vec{x}_k^T - \vec{x}_j \vec{\mu}_2^T - \vec{x}_k \vec{\mu}_2^T  + \vec{\mu}_2 \vec{\mu}_2^T \right] \right\} \\ \nonumber & = Tr\left\{ \Sigma_2^{-1} \left[ \Sigma_1^{(jk)} + \vec{\mu}_1\vec{\mu}_1^T - 2\vec{\mu}_1\vec{\mu}_2^T + \vec{\mu}_2\vec{\mu}_2^T \right] \right\} = Tr \left[ \Sigma_2^{-1} \Sigma_1^{(jk)} \right] + Tr \left[ \Sigma_2^{-1} \left( \vec{\mu}_1 - \vec{\mu}_2 \right) \left( \vec{\mu}_1 - \vec{\mu}_2 \right)^T \right] \\ \nonumber & = Tr \left[ \Sigma_2^{-1} \Sigma_1^{(jk)} \right] + \left( \vec{\mu}_1 - \vec{\mu}_2 \right)^T \Sigma_2^{-1} \left( \vec{\mu}_1 - \vec{\mu}_2 \right)
\end{align} where $\Sigma_1^{(jk)}$ is the cross-covariance matrix between the random vectors $\vec{x}_j$ and $\vec{x}_k$ belonging to the same neighborhood system and computed in the random field with vector of parameters $\vec{\theta}_1 = (\vec{\mu}_1, \Sigma_1, \beta_1)$. Hence, the expression for $B$ becomes:

\begin{align}
	B = \frac{1}{2} \Bigg\{ Tr \left( \Sigma_2^{-1} \Sigma_1 \right) + \left( \vec{\mu}_1 - \vec{\mu}_2 \right)^T \Sigma_2^{-1} \left( \vec{\mu}_1 - \vec{\mu}_2 \right) & - 2\beta_2\sum_{j\in\eta_i} \left[ Tr \left( \Sigma_2^{-1} \Sigma_1^{(ij)} \right) + \left( \vec{\mu}_1 - \vec{\mu}_2 \right)^T \Sigma_2^{-1} \left( \vec{\mu}_1 - \vec{\mu}_2 \right) \right] \\ \nonumber & + \beta_2^2 \sum_{j\in\eta_i} \sum_{k\in\eta_i} \left[ Tr \left( \Sigma_2^{-1} \Sigma_1^{(jk)} \right) + \left( \vec{\mu}_1 - \vec{\mu}_2 \right)^T \Sigma_2^{-1} \left( \vec{\mu}_1 - \vec{\mu}_2 \right) \right] \Bigg\}
\end{align}

After some little algebra, we finally have:

\begin{align}
	B & = \frac{1}{2} \Bigg\{ Tr \left( \Sigma_2^{-1} \Sigma_1 \right) - 2\beta_2\sum_{j\in\eta_i} Tr \left( \Sigma_2^{-1} \Sigma_1^{(ij)} \right) + \beta_2^2 \sum_{j\in\eta_i} \sum_{k\in\eta_i} Tr \left( \Sigma_2^{-1} \Sigma_1^{(jk)} \right) + \left( 1 - \Delta\beta_2 \right)^2 \left[ \left( \vec{\mu}_1 - \vec{\mu}_2 \right)^T \Sigma_2^{-1} \left( \vec{\mu}_1 - \vec{\mu}_2 \right) \right] \Bigg\}
\end{align} where $\Delta$ denotes the number of neighbors (for a second-order neighborhood system, $\Delta=8$).

Therefore, the KL-divergence between two multivariate pairwise isotropic GMRF's   $p(\vec{x}_i | \eta_i, \vec{\theta}_1)$ and $q(\vec{x}_i | \eta_i, \vec{\theta}_2)$, for $i=1,2,...,n$, with parameter vectors $\vec{\theta}_1 = (\vec{\mu}_1, \Sigma_1, \beta_1)$ and $\vec{\theta}_2 = (\vec{\mu}_2, \Sigma_2, \beta_2)$ is given by:

\begin{align}
	D_{KL}(p, q) & = \frac{1}{2} log \frac{\lvert \Sigma_2 \rvert}{\lvert \Sigma_1 \rvert} -\frac{1}{2} \left\{ d - 2\beta_1 \sum_{j\in\eta_i} Tr \left[ \Sigma_1^{-1} \Sigma_1^{(ij)} \right] + \beta_1^2 \sum_{j\in\eta_i} \sum_{k\in\eta_i} Tr \left[ \Sigma_1^{-1} \Sigma_1^{(jk)} \right] \right\} \nonumber \\ & + \frac{1}{2} \Bigg\{ Tr \left( \Sigma_2^{-1} \Sigma_1 \right) - 2\beta_2\sum_{j\in\eta_i} Tr \left( \Sigma_2^{-1} \Sigma_1^{(ij)} \right) + \beta_2^2 \sum_{j\in\eta_i} \sum_{k\in\eta_i} Tr \left( \Sigma_2^{-1} \Sigma_1^{(jk)} \right) \\ \nonumber & \qquad\qquad\qquad\qquad\qquad + \left( 1 - \Delta\beta_2 \right)^2 \left[ \left( \vec{\mu}_1 - \vec{\mu}_2 \right)^T \Sigma_2^{-1} \left( \vec{\mu}_1 - \vec{\mu}_2 \right) \right] \Bigg\}
\end{align}

As the KL-divergence is not symmetric, $D_{KL}(q, p)$ is given by:

\begin{align}
	D_{KL}(q, p) & = \frac{1}{2} log \frac{\lvert \Sigma_1 \rvert}{\lvert \Sigma_2 \rvert} -\frac{1}{2} \left\{ d - 2\beta_2 \sum_{j\in\eta_i} Tr \left[ \Sigma_2^{-1} \Sigma_2^{(ij)} \right] + \beta_2^2 \sum_{j\in\eta_i} \sum_{k\in\eta_i} Tr \left[ \Sigma_2^{-1} \Sigma_2^{(jk)} \right] \right\} \nonumber \\ & + \frac{1}{2} \Bigg\{ Tr \left( \Sigma_1^{-1} \Sigma_2 \right) - 2\beta_1\sum_{j\in\eta_i} Tr \left( \Sigma_1^{-1} \Sigma_2^{(ij)} \right) + \beta_1^2 \sum_{j\in\eta_i} \sum_{k\in\eta_i} Tr \left( \Sigma_1^{-1} \Sigma_2^{(jk)} \right) \\ \nonumber & \qquad\qquad\qquad\qquad\qquad + \left( 1 - \Delta\beta_1 \right)^2 \left[ \left( \vec{\mu}_1 - \vec{\mu}_2 \right)^T \Sigma_1^{-1} \left( \vec{\mu}_1 - \vec{\mu}_2 \right) \right] \Bigg\}
\end{align}

Given the expressions for $D_{KL}(p, q)$ and $D_{KL}(q, p)$, we can compute the symmetrized KL-divergence by their average and use it as a distance function in several image processing and machine learning applications, such as image denoising and unsupervised metric learning.

\subsection{Maximum pseudo-likelihood estimation of the inverse temperature}

As the KL-divergence is a function of the model parameters, we need to know how to estimate the inverse temperature. The log pseudo-likelihood function is:

\begin{align}
	log~L(\vec{\theta}; X) & = - \frac{d}{2} log 2\pi - \frac{1}{2} log \lvert \Sigma \rvert  - \frac{1}{2}\sum_{i=1}^n \left\{ \left[ \vec{x}_i - \vec{\mu} - \beta \sum_{j \in \eta} \left( \vec{x}_j - \vec{\mu} \right)\right]^T \Sigma^{-1} \left[ \vec{x}_i - \vec{\mu} - \beta \sum_{j \in \eta} \left( \vec{x}_j - \vec{\mu} \right)\right]\right\}
\end{align}

Note that the first and second terms do not depend on the inverse temperature, so differentiating with respect to $\beta$ and setting the result to zero leads to:

\begin{align}
	\sum_{i=1}^n \left\{ \left[ \vec{x}_i - \vec{\mu} - \beta \sum_{j \in \eta} \left( \vec{x}_j - \vec{\mu} \right)\right]^T \Sigma^{-1} \left[ \vec{x}_i - \vec{\mu} - \beta \sum_{j \in \eta} \left( \vec{x}_j - \vec{\mu} \right)\right]\right\} = 0
\end{align}

By properly solving the equation, we finally reach a closed-form expression for the maximum pseudo-likelihood estimator of the inverse temperature parameter as:

\begin{align}
	\hat{\beta}_{MPL} = \frac{ \displaystyle \sum_{i=1}^{n} \left[ \sum_{j\in\eta_i} \left( \vec{x}_i - \vec{\mu} \right)^T \left( \vec{x}_j - \vec{\mu} \right) \right] }{ \displaystyle \sum_{i=1}^{n} \left[ \sum_{j\in\eta_i} \sum_{j\in\eta_i} \left( \vec{x}_j - \vec{\mu} \right)^T \left( \vec{x}_k - \vec{\mu} \right) \right] }
	\label{eq:MPL}
\end{align}

It is worth to mention that to have $\beta = 0$, the numerator of equation \eqref{eq:MPL} must be zero. Let us assume, without loss of generality, that the mean vector is zero. The only case in which this situation happens is when $\vec{x}_i$ is orthogonal to all its neighbors $\vec{x}_j \in \eta_i$, for $i=1, 2,... ,n$, which defines a rather rare geometric setting. 

\section{Conclusions and final remarks}

Being able to measure the similarity between random variables is a fundamental task in many area of science, from classical information theory to modern physics applications. In complex systems modeling, for instance, often, it is not clear how to measure the similarity between two systems operating in different regimes. Usually, after a feature extraction stage, where several properties of the systems are obtained and organized into a feature vector, a similarity measure such as the Euclidean distance is employed to quantify how far these vectors are in the feature space. 

In this paper, we derived closed-form expressions for the Kullback-Leibler divergence between univariate and multivariate pairwise isotropic Gaussian-Markov random fields as an attempt to provide an objective way to measure the distance between random fields composed by Gaussian random variables. The motivation was the use of the local conditional density functions, which do not depend on a global partition function. Despite not being symmetric (but being easily symmetrized), one possible advantage of the KL-divergence over other measures is that it is directly related to the Fisher-Rao metric, which allows the computation of intrinsic measures, such as the geodesic distances.

Overall, the proposed closed-form equations are quite intuitive and in case of independent random variables ($\beta = 0$), they are simplified to the KL-divergence between regular Gaussian distributions. A positive aspect is that no numerical algorithm is required in the exact computation of this divergence between pairwise isotropic GMRF's, making it a computationally feasible even when the number of samples grows arbitrarily. 

Future works may include the development of novel image denoising methods using the Non-Local Means technique, as follows: by modeling each patch of the image as an outcome of a univariate pairwise isotropic Gaussian random field, we can compute the similarity between patches using the derived KL-divergence. We also intend to apply the KL-divergence between multivariate pairwise isotropic GMRF's to perform dimensionality reduction based unsupervised metric learning by extending graph-based algorithms as ISOMAP, LLE, Laplacian Eigenmaps: the idea is to replace the regular Euclidean distance to weight the edges of the KNN graph used to approximate the underlying data manifold.

\section*{Acknowledgments}
This study was financed in part by the Coordenação de Aperfeiçoamento de Pessoal de Nível Superior - Brasil (CAPES) - Finance Code 001.





\bibliographystyle{elsarticle-num}
\bibliography{mybibfile}







\end{document}